\documentclass[12pt]{iopart}

\usepackage{graphicx}

\begin{document}

\title[Self-magnetic flux of current-carrying polygons]{Comparing
a current-carrying circular wire with polygons
of equal perimeter:
Magnetic field versus magnetic flux}

\author{J P Silva and A J Silvestre}

\address{Instituto Superior de Engenharia de Lisboa,
Rua Conselheiro Em\'{\i}dio Navarro,
1950-062 Lisboa, Portugal}

\eads{\mailto{jpsilva@deea.isel.ipl.pt},
\mailto{asilvestre@deq.isel.ipl.pt}}

\begin{abstract}
We compare the magnetic field at the center of and the self-magnetic flux
through a current-carrying circular loop,
with those obtained for current-carrying polygons with the same perimeter.
As the magnetic field diverges at the position of the wires,
we compare the self-fluxes utilizing several
regularization procedures.
The calculation is best performed utilizing the vector potential,
thus highlighting its usefulness in practical applications.
Our analysis answers some of the intuition challenges students face
when they encounter a related simple textbook example.
These results can be applied directly to the determination of
mutual inductances in a variety of situations.
\end{abstract}


\maketitle

\section{\label{sec:intro}Introduction}

\begin{figure}[htb]
\begin{center}
\includegraphics*[height=3.5cm]{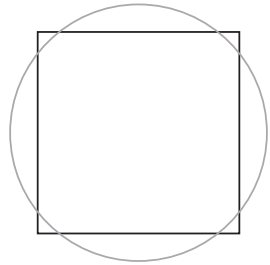}
\caption{\label{figure1}Square and circle of equal perimeter $P$.
}
\end{center}
\end{figure}

A common exercise in introductory physics courses
concerns the comparison between the magnetic fields due to two
loops of equal length $P$,
carrying the same current $i$,
one shaped into a square and the other shaped into a
circle.
One is asked to compare the magnetic fields
at the centers of the respective figures \cite{HRW},
finding that the field at the center of the square is
larger than the field at the center of the circle.
In our classes,
this problem is always followed by a lively debate.
Many students feel that the opposite should occur,
citing the fact that,
for a given perimeter $P$,
the circle is the figure with the largest area.
It is only when the two figures are drawn to scale,
as in figure~\ref{figure1},
that they understand the result.
The point is that,
for equal perimeter,
the sides of the square lie inside the circle for most
of the integration paths.

\begin{figure}[htb]
\begin{center}
\includegraphics*[height=6.5cm]{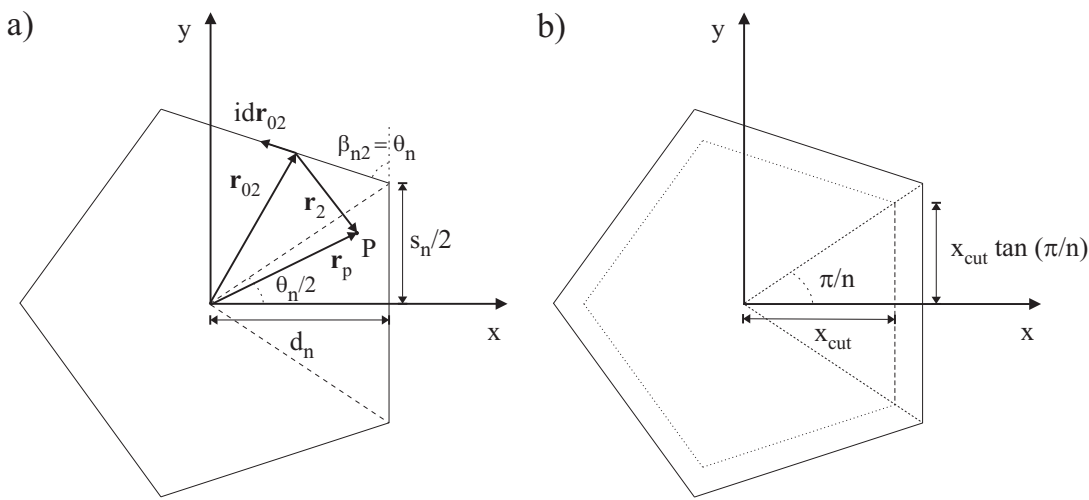}
\caption{\label{figure2}Pentagon with perimeter $P$.
(a) Pictorial representation of the vectors used in the calculation of
$\bi{A}$,
which are defined in the text.
(b) The line integral of $\bi{A}$ is taken along the inner (dotted)
polygonal curve $C_n$.
}
\end{center}
\end{figure}

The result can be easily generalized
for any polygon with $n$ equal sides and total
perimeter $P$.
figure~\ref{figure2},
illustrates the case of $n=5$.
Each side has length $s_n=P/n$,
placed at a distance $d_n = s_n/2\, \cot{(\theta_n/2)}$
from the center,
where $\theta_n=2\pi/n$.
The total  magnetic field is simply equal
to $n$ times the field produced by a straight wire of length
$s_n$ carrying a current $i$,
at a point placed at a distance $d_n$ from the wire,
along its perpendicular bisector:
\begin{equation}
B_n^{\rm center} = n\,
\frac{\mu_0 i}{4 \pi d_n}
\frac{s_n}{\sqrt{(s_n/2)^2 + d_n^2}}
=
\frac{\mu_0 i}{4 \pi P}
4 n^2\,
\tan{(\pi/n)} \sin{(\pi/n)}.
\label{Bncenter}
\end{equation}
Substituting for $n=3, 4, \dots$ in equation~(\ref{Bncenter}),
we conclude that,
for equal perimeter,
the field at the center of a current-carrying triangle is the largest;
and the fields at the center of other current-carrying
polygons with equal perimeter decrease as the number of sides increases,
approaching the asymptotic value of
$B_c^{\rm center}=\frac{\mu_0 i}{4 \pi P} 4 \pi^2$
obtained for the circle.
This calculation can be assigned as a homework exercise.

Although the area does not play a role in this example,
our students usually point out that it should play a role
in determining the auto-flux through the wire loops.
For a given perimeter $P$,
the areas enclosed by the polygon wires are
${\cal A}_n = P^2 \cot{(\pi/n)}/(4n)$,
approaching the area of the circle,
${\cal A}_c = P^2 /(4 \pi)$,
as the number of sides increases.
The naive multiplication
\begin{equation}
B_n^{\rm center} {\cal A}_n =
\frac{\mu_0 i P}{4 \pi}
n \sin{(\pi/n)},
\label{naive}
\end{equation}
grows with $n$.
Normalizing this type of ``flux'' by
$\frac{\mu_0 i P}{4 \pi}$,
as we shall henceforth do\footnote{All
our figures will be drawn for the flux $\Phi$
in units of $\mu_0 i P/(4 \pi)$,
i.e.,
whenever we mention $\Phi$ on the vertical axis,
we are really plotting $4 \pi \Phi/(\mu_0 i P)$.
},
we find $2.6$, $2.8$, $3.1$, and $\pi$
for $n=3$, $n=4$, $n=8$, and the circle,
respectively.
This seems to indicate that the
smaller field at the center of the circle
is more than compensated by its larger area.
Some students interpret this as a vindication of
their initial intuition.

Unfortunately,
things are considerably more complicated than this simple
argument suggests,
making it interesting to revisit this problem in an
advanced course on electromagnetism.
Firstly,
the magnetic field varies from point to point
in space.
The calculations of these magnetic fields
may be found elsewhere
for the polygon \cite{GJ},
for the circular loop \cite{E},
and for planar wires \cite{M}.
Secondly,
these fields diverge at the position of the wires,
meaning that some regularization must be used.
Thirdly,
obtaining the flux directly from the magnetic fields requires
a two dimensional integration,
which becomes particularly difficult in the case of polygons.

In this article,
we start by calculating the vector potential $\bi{A}$ produced
by a circular or polygonal loop of perimeter $P$ and carrying a
current $i$,
at any point in the plane of the figure, inside the figure.
Naturally,
$\bi{A}$ and $\bi{B} = \mathbf{\nabla} \times \bi{A}$ diverge
as one approaches the wire loop.
So,
we will consider the flux of $\bi{B}$ through a
surface $S$ with edges on a curve $C$
similar to (and concentric with) the current loop,
but scaled down by some amount (\textit{c.f.\/} figure~\ref{figure2}(b)).
Obtaining the flux directly from $\bi{B}$ will require a further
two-dimensional integration (besides the one needed to obtain $\bi{B}$),
which, moreover, is rather cumbersome in the case of polygonal surfaces.
Fortunately,
we may use Stokes theorem
\begin{equation}
\int_S \bi{B} \cdot  \rmd\bi{a} =
\int_C \bi{A} \cdot \rmd\bi{l}
\label{AnotB}
\end{equation}
to turn the two-dimensional integration involving
$\bi{B}$ into the one-dimensional integration
involving $\bi{A}$.
Many textbooks only mention the vector potential
briefly;
this problem provides a striking example of how useful
the vector potential may be in practical applications.

The results we obtain also provide the mutual inductance of two nested,
coplanar, and concentric (polygonal or circular) wires of equal
shape but different scales.
This can be used for theoretical discussions and
experimental studies of Faraday's law.

\section{\label{sec:vector_potential}Calculating the vector potential}

We wish to calculate $\bi{A}_n(x,y)$ at a point P with
coordinates
$\bi{r}_P = x\, \mathbf{\hat{e}}_x + y\, \mathbf{\hat{e}}_y$,
as illustrated in figure~\ref{figure2}(a).
We start by parametrizing the positions of
the points on the right-hand side of the polygon as
$\bi{r}_{01} = d_n\, \mathbf{\hat{e}}_x + t\, \mathbf{\hat{e}}_y$,
with $t \in (-s_n/2, s_n/2)$.
Using $\bi{r}_1 = \bi{r}_P - \bi{r}_{01}$,
we find
\begin{eqnarray}
\frac{4 \pi}{\mu_0 i}
\bi{A}_{n1}
&=&
\int_{-s_n/2}^{s_n/2}
\frac{1}{r_1} \frac{\rmd\bi{r}_{01}}{\rmd t} \rmd t
=
\int_{-s_n/2}^{s_n/2}
\frac{\rmd t}{\sqrt{(x-d_n)^2+(y-t)^2}}\ \mathbf{\hat{e}}_y
\nonumber\\*[3mm]
&=&
\ln{
\left\{
\frac{- y + s_n/2
+ \sqrt{\left[x-d_n\right]^2+\left[y-s_n/2\right]^2}
}{- y - s_n/2
+ \sqrt{\left[x-d_n\right]^2+\left[y+s_n/2\right]^2}
}
\right\}
}
\ \mathbf{\hat{e}}_y\ .
\label{An1}
\end{eqnarray}
The position of the points along the $k$-th side
(moving anti-clockwise) is simply given by a rotation
of $\bi{r}_{01}$ by an angle $\beta_{nk} = (k-1) \theta_n = 2\pi(k-1)/n$.
So,
$\bi{r}_{0k} = X_{nk}(t)\, \mathbf{\hat{e}}_x +
Y_{nk}(t)\, \mathbf{\hat{e}}_y$,
where
\begin{eqnarray}
X_{nk}(t) &=& d_n \cos{\beta_{nk}} - t \sin{\beta_{nk}},
\nonumber\\
Y_{nk}(t) &=& d_n \sin{\beta_{nk}} + t \cos{\beta_{nk}}.
\end{eqnarray}
As a result
\begin{eqnarray}
\fl
\frac{4 \pi}{\mu_0 i}
\bi{A}_{nk}
&=&
\int_{-s_n/2}^{s_n/2}
\frac{\rmd t}{\sqrt{\left[x-X_{nk}(t)\right]^2+\left[y-Y_{nk}(t)\right]^2}}
\ \mathbf{\hat{e}}_{nk}
\nonumber\\*[3mm]
\fl
&=&
\ln{
\left\{
\frac{s_n/2 - a_{nk}(x,y)
+ \sqrt{\left[x-X_{nk}(s_n/2)\right]^2+\left[y-Y_{nk}(s_n/2)\right]^2}
}{-s_n/2 - a_{nk}(x,y)
+ \sqrt{\left[x-X_{nk}(-s_n/2)\right]^2+\left[y-Y_{nk}(-s_n/2)\right]^2}
}
\right\}
}
\ \mathbf{\hat{e}}_{nk}
\label{Ank}
\end{eqnarray}
where
\begin{equation}
\mathbf{\hat{e}}_{nk}
=
- \sin{\beta_{nk}}\, \mathbf{\hat{e}}_x +
\cos{\beta_{nk}}\, \mathbf{\hat{e}}_y
\end{equation}
and
\begin{equation}
\pm s_n/2 - a_{nk}(x,y)
=
\left[x-X_{nk}(\pm s_n/2)\right] \sin{\beta_{nk}}
- \left[y-Y_{nk}(\pm s_n/2)\right] \cos{\beta_{nk}}.
\label{a_nk}
\end{equation}
The final magnetic vector potential is given by
\begin{equation}
\bi{A}_n(x,y)
=
\sum_{k=1}^{n}
\bi{A}_{nk}(x,y).
\label{Antotal}
\end{equation}
Alternatively,
we might obtain equation~(\ref{Ank}) from equation~(\ref{An1})
through the vector field
rotations discussed by Grivich and Jackson \cite{GJ}.
We could now recover their Equation (9) with $z=0$
by taking $\bi{B} = \mathbf{\nabla} \times \bi{A}$
and suitable variable redefinitions\footnote{There
is a subtlety concerning the fact that,
since we have determined $\bi{A}(x,y,z)$ only for the plane $z=0$,
we cannot perform the derivations with respect to $z$.
However,
these do not enter the calculation of $B_z(x,y,0)$ which,
by symmetry,
is the only non-vanishing component
of $\bi{B}(x,y,z)$ when $z=0$.
}.

As for the circular loop,
we use polar coordinates.
By symmetry,
\begin{equation}
\bi{A}_c(\rho,\theta)
=
A_c(\rho,\theta)\, \mathbf{\hat{e}}_{\theta}
=
A_c(\rho,0)\, \mathbf{\hat{e}}_{\theta},
\label{Ac_not_theta}
\end{equation}
and we take $\bi{r}_P = \rho\, \mathbf{\hat{e}}_x$.
Parametrizing the positions of
the points along the current-carrying circular wire
of radius $R$ as
$\bi{r}_{0} = R \cos{\varphi}\, \mathbf{\hat{e}}_x
+ R \sin{\varphi}\, \mathbf{\hat{e}}_y$,
with $\varphi \in (0, 2\pi)$,
$\bi{r} = \bi{r}_P - \bi{r}_{0}$,
and we find
\begin{eqnarray}
\fl
\frac{4 \pi}{\mu_0 i}
\bi{A}_c(\rho,0)
&=&
\int_{0}^{2\pi}
\frac{1}{r} \frac{\rmd\bi{r}_0}{\rmd\varphi} \rmd\varphi
=
\int_{0}^{2\pi}
\frac{- R \sin{\varphi}\, \mathbf{\hat{e}}_x
+ R \cos{\varphi}\, \mathbf{\hat{e}}_y
}{\sqrt{\rho^2 + R^2 - 2 \rho R \cos{\varphi}}}\ \rmd\varphi
\nonumber\\*[3mm]
\fl
&=&
\frac{2}{\rho(\rho+R)}
\left[
(\rho^2+R^2)\, K\left(\frac{2\sqrt{\rho R}}{\rho+R}\right)
-(\rho+R)^2\, E\left(\frac{2\sqrt{\rho R}}{\rho+R}\right)
\right]\, \mathbf{\hat{e}}_y\ ,
\label{Ac}
\end{eqnarray}
where
\begin{equation}
K(k) =
\int_0^1 \frac{\rmd t}{\sqrt{1-k^2 t^2} \sqrt{1-t^2}},
\ \ \
E(k) =
\int_0^1 \frac{\sqrt{1-k^2 t^2}}{\sqrt{1-t^2}} \rmd t.
\end{equation}
We have checked that the function
$A_n(\rho,0)$ in equation~(\ref{Antotal}) tends to
$A_c(\rho,0)$ in equation~(\ref{Ac}),
as $n$ approaches infinity.
Also,
by taking $\bi{B} = \mathbf{\nabla} \times \bi{A}$
and suitable variable redefinitions,
we recover the corresponding magnetic field \cite{E}.

\section{\label{sec:flux}Calculating the flux}

We recall two points mentioned in the introduction.
Because the fields diverge at the position of the wires,
we will take the flux in a curve similar to the original wire but
scaled down by some amount,
as in figure~\ref{figure2}(b).
We may think of this as a cutoff introduced by the finite width of
the wire,
or as the situation faced in calculating the flux through a second
loop,
similar to (but smaller than) the current-carrying one.
Also,
because the direct calculation of the flux of $\bi{B}$ involves
a two-dimensional integration,
we will use equation~(\ref{AnotB}) and calculate instead
the line integral of $\bi{A}$.

The simplicity gained in utilizing $\bi{A}$
is particularly striking in the case of the circular current loop,
since equation~(\ref{Ac_not_theta}) means that $\bi{A}$ is
independent of $\theta$.
Therefore,
choosing an integration circle $C_\rho,$ of radius
$\rho \in (0,R)$,
we find
\begin{eqnarray}
\frac{4 \pi}{\mu_0 i P}
\Phi_c
&=&
\frac{4 \pi}{\mu_0 i P}
\int_{C_\rho} \bi{A} \cdot \rmd\bi{l}
=
\frac{4 \pi}{\mu_0 i P}
A(\rho,0) \ 2\pi\rho
\nonumber\\*[3mm]
&=&
\frac{4 \pi}{\rho+R}
\left[
(\rho^2+R^2)\, K\left(\frac{2\sqrt{\rho R}}{\rho+R}\right)
-(\rho+R)^2\, E\left(\frac{2\sqrt{\rho R}}{\rho+R}\right)
\right],
\label{Phic}
\end{eqnarray}
where,
in going to the second line,
we have made $\rho$ and $R$ dimensionless by scaling them
by the perimeter $P$\footnote{We
have made the variable substitutions $\rho^\prime = \rho/P$ and
$R^\prime = R/P = 1/(2\pi)$, and then dropped the primes.
}.
It is instructive to compare the trivial reasoning on the first line of
equation~(\ref{Phic}) with what would be needed to calculate
the flux directly from the $\rho$-dependent
$\bi{B}_c$.

Next we consider the magnetic field produced by a polygon with
perimeter $P$,
$n$ equal sides,
and carrying the current $i$.
The distance from the center to each of the sides is given
by $d_n$.
Consider also a second $n$-sided polygon $C_n$ whose sides lie a
distance $x_{\rm cut} \in (0,d_n)$ from the same center.
The flux through this polygon is given by
\begin{equation}
\Phi_n =
\int_{C_n} \bi{A}_n \cdot \rmd\bi{l}
=
n \int_{\rm first\ side} \bi{A}_n \cdot \rmd\bi{l}
=
n \int_{-x_{\rm cut} \tan{(\pi/n)}}^{x_{\rm cut} \tan{(\pi/n)}}
{(A_n)}_y(x_{\rm cut},y)\,  \rmd y.
\label{Phin_setup}
\end{equation}
Looking back at equation~(\ref{Ank}) one notices the need for integrals
involving the logarithm of rather complicated functions.
Things can be greatly simplified,
however.
We start by rescaling all distances by the perimeter $P$,
thus rendering the variables $x$, $y$, $s_n$, and $d_n$
appearing in equation~(\ref{Ank}) dimensionless\footnote{We
have made the variable substitutions $x^\prime = x/P$,
$y^\prime = y/P$,
$s_n^\prime = s_n/P = 1/n$,
and $d_n^\prime = d_n/P = \cot{(\theta_n/2)}/(2n)$,
and then dropped the primes.
}.
Next we introduce new parameters $u$ and new variables $v$
through
\begin{eqnarray}
u &=&
x_{\rm cut} - X_{nk}(\pm s_n/2),
\nonumber\\
v &=&
y - Y_{nk}(\pm s_n/2),
\end{eqnarray}
for use in equations~(\ref{Ank}) and (\ref{a_nk}).
Thus,
for equation~(\ref{Ank})
we need
\begin{equation}
I_{nk}(u,v) \equiv
\int
\ln{
\left[
u \sin{\beta_{nk}} - v \cos{\beta_{nk}}
+ \sqrt{u^2+v^2}
\right]
}\ \rmd v\, .
\end{equation}
We find\footnote{We are very grateful to Ana C. Barroso for help with
this integral.}
\begin{equation}
\fl
I_{nk}[u,v]
=
\left\{
\begin{array}{ll}
     v\, \ln{\left(-v+\sqrt{u^2+v^2}\right)}+\sqrt{u^2+v^2} &
                          \mbox{\ if\ \ } \beta_{nk}=0\\*[3mm]
     v\, \ln{\left(v+\sqrt{u^2+v^2}\right)}-\sqrt{u^2+v^2} &
                          \mbox{\ if\ \ } \beta_{nk}=\pi\\*[3mm]
     - v + u\, \csc{\beta_{nk}}\, \ln{\left(v+\sqrt{u^2+v^2}\right)} & \\
          \hspace{4ex} +\, (v + u \cot{\beta_{nk}})
                \ln{\left(u \sin{\beta_{nk}} - v \cos{\beta_{nk}}
                                          + \sqrt{u^2+v^2}\right)} &
                   \mbox{\ otherwise}\, .
\end{array}
\right.
\end{equation}
Combining this with equations~(\ref{Ank})--(\ref{Antotal}),
and
substituting into equation~(\ref{Phin_setup}),
one obtains
\begin{equation}
\frac{4 \pi}{\mu_0 i P}
\Phi_n
=
n\, \sum_{k=1}^n \cos{\beta_{nk}}
\left(
I^+_{nk} - I^-_{nk}
\right),
\label{Phin}
\end{equation}
where
\begin{eqnarray}
I^\pm_{nk} &=&
I\left[
x_{\rm cut} - X_{nk}(\pm s_n/2),
\ x_{\rm cut} \tan{(\pi/n)} - Y_{nk}(\pm s_n/2)
\right]
\nonumber\\
& &
-
I\left[
x_{\rm cut} - X_{nk}(\pm s_n/2),
\, - x_{\rm cut} \tan{(\pi/n)} - Y_{nk}(\pm s_n/2)
\right].
\end{eqnarray}

\begin{figure}[htb]
\begin{center}
\includegraphics*[height=7cm]{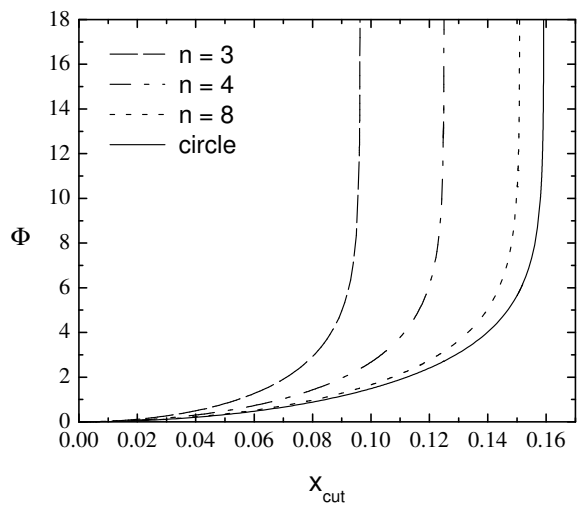}
\caption{\label{figure3}Auto-fluxes as a function of
$x_{\rm cut}$,
for current-carrying polygons with  $n=3$, $4$, $8$,
and for the circular loop.
}
\end{center}
\end{figure}

We have checked equations~(\ref{Phic}) and (\ref{Phin})
in two important limits.
First, expanding around $x_{\rm cut}=0$,
we find that the fluxes tend to the product of the
magnetic field at the center with the area of a small
central region whose distance to the sides is $x_{\rm cut}$.
Indeed,
\begin{eqnarray}
\fl
\Phi_c
& \rightarrow &
B_c^{\rm center}\ \pi\, x_{\rm cut}^2 + O(x_{\rm cut}^3)
\ \rightarrow\
4 \pi^3\ x_{\rm cut}^2\, ,
\label{limit1-1}
\\
\fl
\Phi_n
 & \rightarrow &
B_n^{\rm center}\ n \tan{(\pi/n)}\, x_{\rm cut}^2 + O(x_{\rm cut}^3)
\ \rightarrow\
4 n^3 \tan^2{(\pi/n)} \sin{(\pi/n)}\ x_{\rm cut}^2\, .
\label{limit1-2}
\end{eqnarray}
Here and henceforth (including in all figures),
we normalize the fluxes by $\mu_0 i P/(4 \pi)$,
the magnetic fields by $\mu_0 i/(4 \pi P)$,
and we continue to scale all distances by $P$.
Naturally,
we can recover equation~(\ref{limit1-1}) from equation~(\ref{limit1-2}) in the
limit of $n$ going to infinity.
Second,
$\Phi_n$ tends to $\Phi_c$ as $n$ goes to infinity,
for all values of $x_{\rm cut}$.
This can be seen in figure~\ref{figure3},
which displays $\Phi_n$ for $n=3$, $4$, $8$,
and $\Phi_c$ as a function of $x_{\rm cut}$.
Each flux $\Phi_n$ diverges at $x_{\rm cut} = d_n$,
while $\Phi_c$ diverges at $x_{\rm cut} = R$,
providing a nontrivial crosscheck on our expressions.
Notice that,
for each value of $x_{\rm cut} < d_3$,
the curve for $\Phi_c$ lies below all other fluxes.
Although the fields $\bi{B}$ vary as one moves away
from the center,
a very rough way of understanding this result
is the following:
the field at the center $B_n^{\rm center}$ decreases as
$n$ increases---\textit{c.f.\/} equation~(\ref{Bncenter});
on the other hand,
for fixed $x_{\rm cut}$,
the areas through which the flux is being considered
are given by $n \tan{(\pi/n)}\ x_{\rm cut}^2$,
for $\Phi_n$,
and by $\pi\ x_{\rm cut}^2$,
for $\Phi_c$,
which also decrease as $n$ increases.
Therefore,
in this case the ``area factors'' do not compensate for
the smaller fields,
as seen in equations~(\ref{limit1-1}) and (\ref{limit1-2}).

\begin{figure}[htb]
\begin{center}
\includegraphics*[height=7cm]{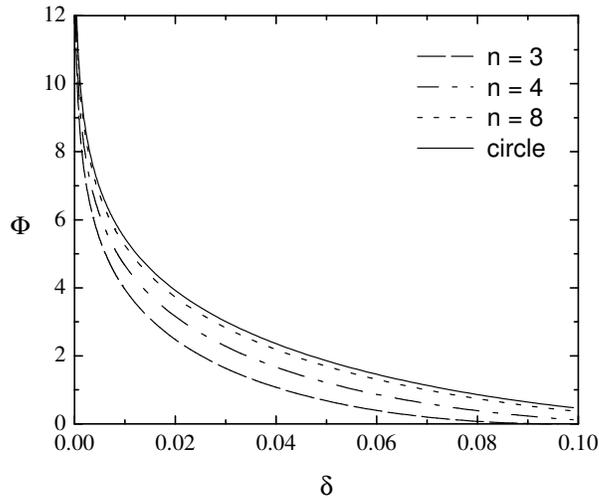}
\caption{\label{figure4}Auto-fluxes as a function of
the width of the wire,
for current-carrying polygons with  $n=3$, $4$, $8$,
and for the circular loop.
}
\end{center}
\end{figure}

Since the fluxes diverge for $x_{\rm cut}=d_n$,
we may choose to consider another situation.
We take all wires to be of a fixed width $\delta$ (in units of $P$),
and we regularize the fluxes by integrating only up to
$\rho = R - \delta$, for the circle,
and $x_{\rm cut} = d_n - \delta$,
for the polygons.
The results are displayed in figure~\ref{figure4}
as a function of $\delta$,
for $n=3$, $4$, $8$,
and for the circle.
We notice the following features:
i) for any finite value of $\delta$,
the auto-flux increases as $n$ increases---this
indicates that, here, the ``area factor'' is making up for the
smaller value of the magnetic field at the center;
ii) again, the curves of $\Phi_n$ tend to $\Phi_c$ as
$n$ increases;
iii) the flux diverges as the width of the wires tends to zero,
as expected.

Comparing figure~\ref{figure3} and figure~\ref{figure4}
we notice that $\Phi$ decreases with $n$ in the first case,
while it increases with $n$ in the second.
So,
in contrast to the previous case,
here the ``area factor'' compensates for the smaller
fields.
We can get a rough understanding for this in the following way:
for fixed $\delta$,
the areas through which the flux is being considered
are given by
\begin{equation}
n \tan{(\pi/n)}\, (d_n - \delta)^2
=
\frac{1}{4n} \cot{(\pi/n)} - \delta + n \tan{(\pi/n)}\ \delta^2,
\label{limit2-1}
\end{equation}
for $\Phi_n$,
and by
\begin{equation}
\pi (R - \delta)^2
=
\frac{1}{4 \pi} - \delta + \pi\ \delta^2,
\label{limit2-2}
\end{equation}
for $\Phi_c$,
in units of $P^2$.
As $\delta$ vanishes,
the areas in equations~(\ref{limit2-1}) and (\ref{limit2-2})
are dominated by their first terms,
which do increase enough as to offset the order of the
field magnitudes.
Of course,
this is a very crude argument,
since, because the fields vary in different ways as one
moves away from the center,
using $B_n^{\rm center}$ in the reasoning is a considerable
source of error.
Nevertheless,
this rough argument is consistent with figure~\ref{figure4}.

One can show that,
although the curve of $\Phi_c$ lies above those of $\Phi_n$
for $\delta \neq 0$,
the ratios $\Phi_n/\Phi_c$ tend to one as $\delta$
approaches zero.
This might be difficult to guess initially,
since it seems to contradict the ``area factor'',
but it has an interesting interpretation in terms
of the line integral of
$\bi{A}$.
For points very close to the wires,
the field approaches that of an infinite wire and
$\bi{A}$ diverges logarithmically.
Consequently,
we may interpret the result of the line integral as the product of
a logarithmic divergence with the perimeter $P$ over which
the integral is taken.
Since these features are common to all the current-carrying loops,
all ratios approach unity.
Of course, the same would not be the case if we had taken
current-carrying loops of different perimeter
(recall that we have normalized all fluxes by $\mu_0 i P/(4 \pi)$).

We can choose other regularizations besides the ones discussed above
(constant $x_{\rm cut}$ and constant $\delta$).
For instance,
we may ask that the surfaces through which the flux is being considered
have the same area.
In this case,
as in the case of fixed $x_{\rm cut}$,
$\Phi_n$ decreases as $n$ increases.
In contrast,
if we ask that the surfaces through which the flux is being considered
have the same perimeter,
then $\Phi_n$ increases as $n$ increases,
as in the case of fixed $\delta$.
One can get a rough understanding for these features
along the lines of the analysis made above.

Finally,
we recall that the line integrals of $\bi{A}$ have been
performed over curves $C_n$ and $C_\rho$ identical to
the current-carrying wires, but smaller.
This is what one needs for the calculation of the
mutual inductance between two (polygonal or circular)
current-carrying wires of equal shape and different scales
that lie on the same plane and are concentric.
Our results apply directly to that case.

\section{Conclusions}

Motivated by a simple exercise in elementary electromagnetism,
we have studied the interplay between the magnetic fields
and the areas of current-carrying polygonal and circular
wires of equal perimeter.
We have calculated the vector potential $\bi{A}$
for these situations,
because its line integral provides a much simpler way of
computing the magnetic fluxes;
this example illustrates the usefulness of
$\bi{A}$ in practical calculations.
Since the corresponding auto-fluxes diverge,
we have discussed a number of regularizations,
comparing the fluxes in each case,
and seeking intuitive arguments for the results.
As a bonus,
our results can be applied directly to the calculation of
mutual inductances in a variety of situations.

%

\ack

We are very grateful to Ana C. Barroso for considerable help with some
integrations,
to A. Nunes for reading and commenting on this manuscript,
and to our students for their prodding questions.


\end{document}